\documentclass[aip,reprint]{revtex4-2}

\usepackage{algorithm}
\usepackage{algpseudocode}
\usepackage{lipsum}
\usepackage{amsmath}
\usepackage{amssymb}
\usepackage{graphicx}
\usepackage{gensymb}
\usepackage{tabularx}
\usepackage{float}
\floatstyle{plain}
\restylefloat{table}  
\usepackage{xcolor}
\usepackage{mathtools}
\usepackage{hyperref} 
\usepackage[capitalise]{cleveref}
\DeclareUnicodeCharacter{2009}{ }
\DeclareUnicodeCharacter{2192}{$\rightarrow$}
\DeclareUnicodeCharacter{2032}{'}
\DeclareUnicodeCharacter{03B1}{$\alpha$}
\DeclareUnicodeCharacter{03B3}{$\gamma$}
\restylefloat{table}
\graphicspath{ {images/} }

\begin{document}
	\title{High-dimensional inverse design of inertial fusion implosions via differentiable simulation}
	\author{A. J. Crilly}\email{ac116@ic.ac.uk}
	\affiliation{Centre for Inertial Fusion Studies, The Blackett Laboratory, Imperial College, London SW7 2AZ, United Kingdom}
    \affiliation{I-X Centre for AI In Science, Imperial College London, White City Campus, 84 Wood Lane, London W12 0BZ, United Kingdom}
    \author{P. Travis}
    \affiliation{Ergodic LLC, Seattle, WA 98103}
    \author{J. P. Brodrick}
    \affiliation{Pasteur Labs, Brooklyn, NY}
    \author{J. B. Coughlin}
    \affiliation{Pasteur Labs, Brooklyn, NY}
    \author{A. S. Joglekar}
    \affiliation{Pasteur Labs, Brooklyn, NY}
    \affiliation{Ergodic LLC, Seattle, WA 98103}

    \begin{abstract}
    Inertial confinement fusion implosion design requires simultaneous optimisation of strongly coupled target and driver parameters across high-dimensional design spaces. Existing automated design approaches typically rely on non-differentiable radiation-hydrodynamics codes treated as black boxes, making optimisation increasingly expensive as dimensionality grows. In this work, we present a differentiable simulation approach for high-dimensional inverse design of inertial confinement fusion implosions. Automatic differentiation through a differentiable implosion physics model, driven by an external pressure pulse, provides gradients of implosion objectives with respect to design parameters, enabling gradient-based optimisation. The framework is applied to 25 kJ OMEGA-scale direct-drive implosions, optimising 500-parameter laser pulses across sampled target geometries. The optimised pulse recovers a near-isoentropic rise to peak power without that structure being imposed. Neural-network pulse parameterisations are additionally explored as a means of accelerating design-space exploration. These results establish differentiable implosion modelling as a promising tool for ICF design, while motivating further work on adjoint robustness and higher-fidelity differentiable simulators.
    \end{abstract}

    \maketitle

\section{Introduction}
Inertial confinement fusion (ICF) aims to bring deuterium–tritium (DT) fuel to thermonuclear conditions by using driver energy to compress and heat the fuel via a spherical implosion \cite{Atzeni2004,lindl1995}. In laser direct drive\cite{craxton2015direct} (LDD), a set of laser beams strikes the fuel-filled capsule outer surface directly, producing ablation that acts like a rocket, driving the shell inward. Reaching ignition\cite{Abu2022,Zylstra2022ign,Kritcher2022ign} and sustaining significant burn propagation depend on precisely balancing target geometry with the laser pulse shape to create a fast, but stable, implosion by managing shock timing, adiabat, and implosion trajectory. Because these factors interact strongly, the design space is both high-dimensional and tightly coupled. Additionally, current ICF experiments are expensive and repetition rate is low, and therefore the design of experiments relies heavily on simulation.

Recently, machine learning methods have proven successful in the automated design of ICF implosions\cite{gopalaswamy2025automated,Hatfield2019,Gopalaswamy2021,Ejaz2024,bronner2025particle,Wang2024,Spears2025,Humbird2019,vazirani2021coupling,vazirani2024bayesian,vazirani2025comparison}. These methods iteratively improve designs by launching simulations at varied design parameters, using the results to learn a mapping between design parameters and implosion performance, and therefore find optimal designs. However, the established simulation codes used in these machine learning design processes are treated as a black box, often with surrogate models built around their predictions. This leads to key limitations in the performance of machine learning led design, as the machine learning models have no strong priors on the effect of design choices and therefore new simulations must be run in areas of uncertainty. Additionally, the gradient-free optimisation methods used with the black-box simulation codes often scale poorly with the number of design parameters. Gradient-based optimisation methods present a potential solution to these issues. Firstly, local information on parameter sensitivities/gradients allow new simulations to be run only where one expects an improvement given the parameter gradients. Secondly, gradient-based optimisation scales to very large parameter spaces. However, accurate gradient information is not natively available from existing ICF simulation codes.

In this work, we describe a methodology for obtaining gradient information by applying automatic differentiation methods to a novel implosion code. The code implements canonical numerical solvers for the minimal, reduced physics model of ICF implosions, based on an external pressure drive. The code is written in JAX\cite{jax} enabling the use of automatic differentiation. We show how the derived gradient information can be used for high dimensional optimisation problems and explore the requirements on the adjoint process (used to obtain gradient information) to ensure stable and efficient optimisation. We showcase the inverse design of a high dimensional laser pulse and target geometry for laser direct drive using 25 kJ of laser energy, occupying the design space of the OMEGA experimental facility\cite{Boehly1997}.

\section{Methodology}
\subsection{Automatic Differentiation and Inverse Design}

Automatic differentiation (AD) is a central tool in modern machine learning. It allows the evaluation of partial derivatives of functions within a computer program to machine precision. This is performed by tracing the series of primitive operations performed within the function (its computational graph) and applying the chain rule to autonomously construct the series of operations which evaluates the derivatives of the function. This allows gradients to be efficiently computed for complex functions involving many parameters, the canonical example being neural networks. However, this technique can equally be applied to numerical simulation models, allowing parameter sensitivities to be evaluated following the same operations used to compute the ``forward'' solution in the numerical model. AD can most efficiently compute the gradients of a many-to-one function in reverse, where repeated application of the chain rule is applied from the final operation back through the computational graph. This is also known as back-propagation. 

The back-propagation of gradients through the forward numerical solution to a system of PDEs implicitly solves the adjoint state problem\cite{lions1971}. This approach is known as `discretise-then-optimise'\cite{kidger2022,giles2006} due to the fact that the AD is applied on the forward solution's computational graph, i.e. the series of operations performed to compute the solution to the PDE. As the backpropagation of gradients also requires forward solution values at each step, this leads to large memory requirements of order $\mathcal{O}(N_{\mathrm{temporal}} \times N_{\mathrm{fields}} \times N_{\mathrm{spatial}})$. Checkpointing\cite{gholami2019} is an established strategy for reducing the memory requirements by only saving a reduced number of time-steps. During back-propagation, the required forward solution values are re-computed using the checkpoints to restart from the checkpointed time step. The optimal number of checkpoints represents a balance between memory and computational cost.

\begin{figure}[h]
\centering
\includegraphics*[width=0.99\columnwidth]{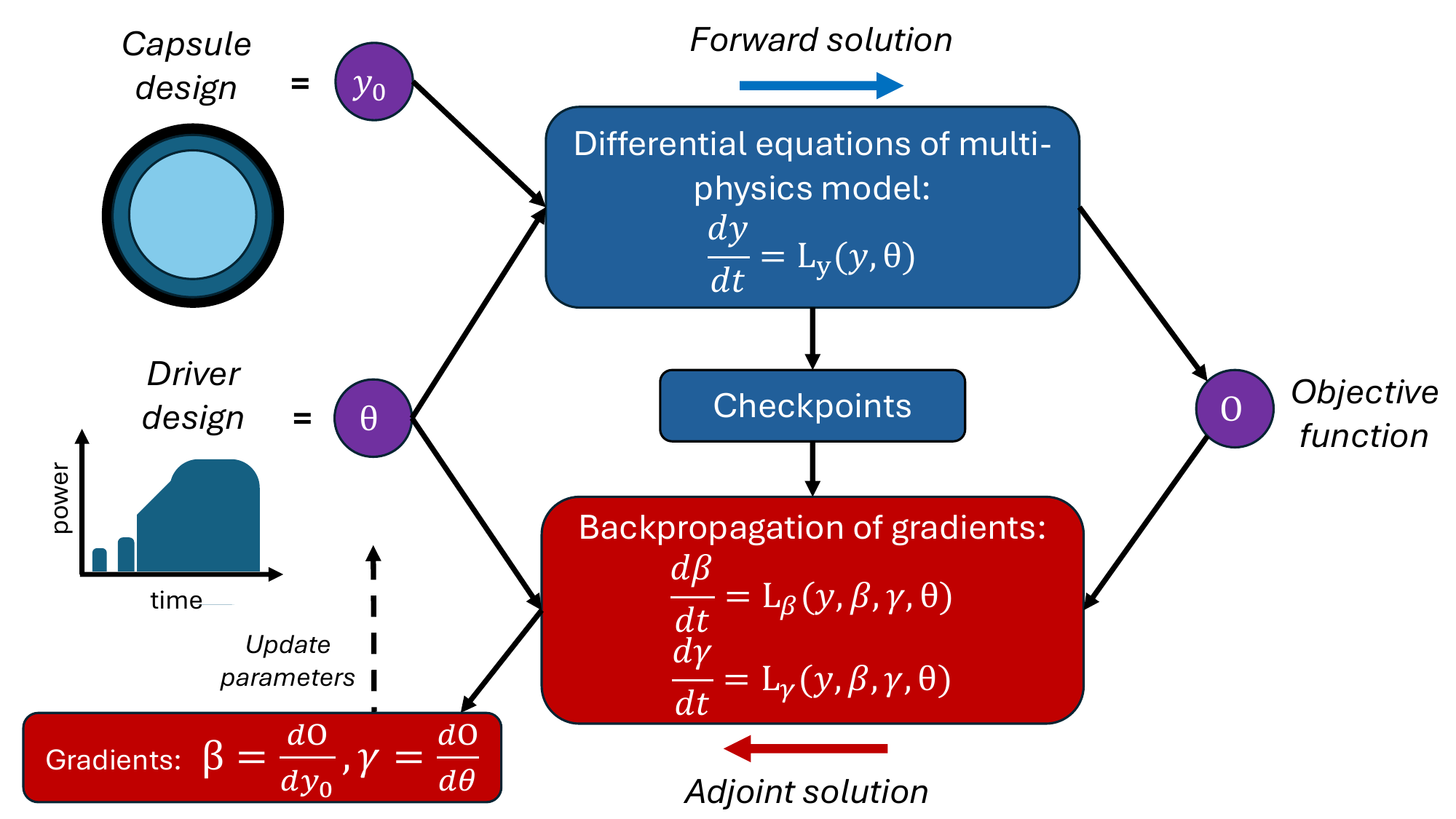}
\caption{Schematic of the differentiable implosion design scheme. Capsule and driver designs are passed to the forward solver as initial conditions and boundary conditions. Checkpoints of the forward solution are saved and the objective function computed. The backpropagation of gradients is performed using automatic differentiation, using the checkpoints to restart the forward solver. Gradients with respect to the objective function are obtained and used to update trainable parameters via gradient descent.}
\label{fig:AD_Diagram}
\end{figure}

Gradient information is particularly useful in an optimisation context. Gradient-based optimisation methods have empirically been shown to scale to very high dimensional spaces involving billions of trainable parameters. Therefore, applying automatic differentiation to numerical solvers to obtain differentiable simulators is an attractive opportunity when considering design problems. \cref{fig:AD_Diagram} shows a diagrammatic representation of the differentiable design method for ICF. For optimisation, we define a scalar loss or objective function which when optimised represents the optimal choice of parameters. One can view the design problem as a constrained optimisation problem, where we optimise our objective subject to the constraint of the physical model (usually taking the form of a system of differential equations). Historically, obtaining gradient information for the objective function constrained by a differential equation model involved the separate derivation of ``adjoint'' equations. However, AD provides a method to derive the required adjoint solution via inspection of the forward numerical solver of the physical system of equations. Therefore, the suitable application of AD to simulators of inertial fusion implosions presents a path to design with a very large number of parameters.

\subsection{Differentiable Implosion Code}

In order to exploit automatic differentiation and implicitly construct the adjoint solution, we must write our numerical model within a differentiable programming framework. In this work, we use JAX\cite{jax} and a number of key libraries (equinox\cite{kidger2021equinox}, lineax, optimistix, optax, diffrax\cite{kidger2022}). These provide a framework within which we must write our numerical solver in a manner which is compatible with AD. One must also decide the minimal set of physical models to investigate implosion design as re-creating the full multi-physics capabilities of state-of-the-art radiation hydrodynamics simulation codes is a considerable task. Therefore, instead of modelling the full laser drive and ablation physics, we follow Herrmann, Tabak, and Lindl (HTL)\cite{HTL2001} and apply an external pressure boundary condition to the fuel in order to drive the implosion. In laser direct drive, some of the fuel is ablated by the laser so here we only consider the unablated fuel which has mass equal to the stagnated mass. This then defines our minimal physical model as the following: compressible Lagrangian hydrodynamics, non-ideal (tabular) equation of state, separate electron and ion energy equations including flux-limited thermal diffusion, and a local radiative loss model based on the DT bremsstrahlung emission.

Given our approach following HTL's external pressure drive, we require a simple model to convert incident laser power into pressure on the fuel. Indeed, an incident laser intensity can be converted to the corresponding ablation pressure via:
\begin{equation}\label{eqn:pressuredrive}
    P_{abl.} = 40 \ \mathrm{Mbar} \left(\frac{I_{15}}{\lambda_{um}}\right)^{\frac{2}{3}}
\end{equation}
where $I_{15}$ is the laser intensity in units of $10^{15} W/cm^2$ and $\lambda_{um}$ is the laser wavelength in microns. A shortcoming of this approach is that we assume pressure is instantly transmitted from the ablation front to the imploding mass and that the boundary of the imploding mass has no inertia. A result of this is that once the laser power is shut off, rarefaction of the fuel begins instantaneously. This will bias the derived laser pulses such that some correction will be required to use these pulses on the full integrated problem.

The forward solution makes use of canonical numerical methods for the required physics. For the compressible Lagrangian hydrodynamics, we follow the staggered grid method of von Neumann and Richtmyer (VNR)\cite{vonneumann1950}. Equation of state tables are produced by the Frankfurt Equation of State (FEOS)\cite{FEOS1} model, which is an extension of QEOS\cite{QEOS}. Root-finding and linear interpolation on these tables were implemented in a fully-differentiable manner. Electron and ion transport coefficients follow Epperlein and Haines\cite{epperlein1986} and NRL plasma formulary\cite{nrl}. Flux-limited heat transport is solved semi-implicitly with operator splitting of equilibration and diffusion, leading to separate tridiagonal equations, solved using the lineax library.

A requirement to enable accurate and stable adjoint solutions is that the forward solvers uses continuously differentiable functions. A key example of this is in the VNR artificial viscosity. In the original formulation, the viscous pressure involves a conditional on the divergence of velocity. An improvement in adjoint solution stability was found with the following modification:
\begin{equation}\label{eqn:smoothedQ}
    \mathrm{min}(0,\Delta u) \rightarrow \frac{1}{2}\left(\Delta u - \sqrt{\Delta u^2 + \epsilon}\right) \ ,
\end{equation}
where $\epsilon$ is a suitably small number (e.g. $10^{-20}$). In \cref{sodshock} we apply AD-derived adjoint solutions to the Sod shock compressible hydrodynamics problem. This simplified problem will be used to illustrate the differentiable inverse design methodology before applying it to laser direct drive implosions in \cref{LDD_design}.

\section{Inverse design in the Sod Shock Tube}\label{sodshock}

The Sod shock tube is a canonical compressible hydrodynamics problem where an initially discontinuous jump in hydrodynamic conditions is evolved in time, with an analytic solution available. We will use this test problem to construct an inverse design problem as an example of the adjoint solution method, showcasing the novel differentiability of our implosion code.

To proceed, we need to define the following:
\begin{itemize}
    \item An objective/loss function to be minimised
    \item A parameterisation of the initial conditions
    \item A numerical scheme for solving the forward and adjoint problems
\end{itemize}
In total, we aim to learn the initial conditions from minimising the mean squared distance to the analytic, reference solution evaluated at a later time. More explicitly, we define a loss function with error and regularisation terms:
\begin{align}
    L(\theta_{\rho},\theta_P) &= MSE(\theta_{\rho},\theta_P) + R(\theta_{\rho},\theta_P) \ , \\
    MSE(\theta_{\rho},\theta_P) &= \frac{1}{N_{\mathrm{zones}}}\sum_{i \in \mathrm{zones}} \frac{1}{\rho_0^2}(\rho^N_i - \rho_{ref, i})^2 \\
    &+ \frac{1}{P_0^2}(P^N_i - P_{ref,i})^2  \ , \nonumber \\
    R(\theta_{\rho},\theta_P) &= \frac{\lambda}{N_{\mathrm{zones}}} \sum_{i \in \mathrm{zones}} \left(\rho_{i}^0 - \rho_{i+1}^0\right)^2 \\
    & + \left(P_{i}^0 - P_{i+1}^0\right)^2   \ , \nonumber
\end{align}
where the superscript is time step and subscript spatial zone, reference data for the terminal solution comes from ExactPack\cite{exactpack} and $\lambda$ is a hyperparameter controlling the degree of regularisation. The initial conditions parameterisation converts the trainable parameters $(\theta_{\rho},\theta_P)$ to bounded initial conditions:
\begin{align}
    \rho^0(\theta_\rho) &= \rho_0 f(\theta_\rho) \ , \\
    P^0(\theta_P) &= P_0 f(\theta_P) \ , \\
    f(\theta) &= 0.01 + \frac{1.99}{1+e^{-\theta}} \ .
\end{align}
The sigmoid transformation ensures the solution remains bounded which aids the optimisation process and prevents unphysical values, i.e. negative pressure and density. A total of 450 spatial Lagrangian zones were used, leading to 900 trainable parameters. At the start of optimisation, all trainable parameters $(\theta_{\rho},\theta_P)$ values were set to zero.

\begin{figure}[h]
\centering
\includegraphics*[width=0.95\columnwidth]{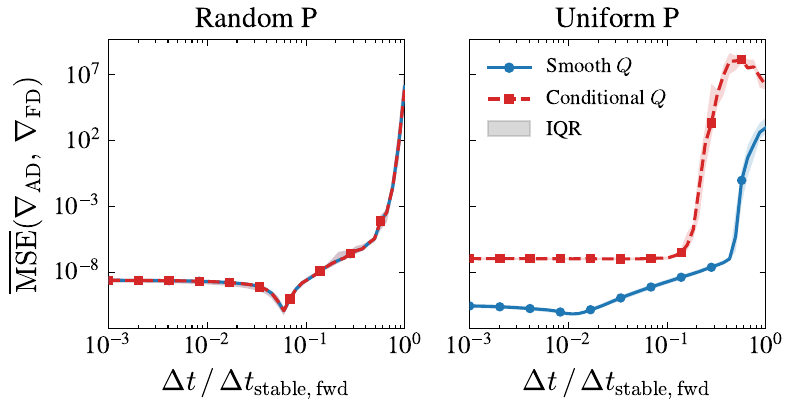}
\caption{Plots showing the mean squared distance between gradients computed using automatic differentiation (AD) and finite difference. Finite difference gradients were computed at a fixed timestep and fixed finite difference step ($\epsilon$ = 1e-8) while the time step for the AD-derived gradients was varied. (Left) Results for various random perturbations of initial conditions. The dip in MSE occurs at the timestep at which the finite difference gradients are computed. (Right) Results for random density but uniform pressure initial conditions. Note that there will be no time evolution for these conditions but still non-zero gradients with respect to initial conditions.}
\label{fig:gradient_stability_analysis}
\end{figure}

The numerical scheme of the forward solution was staggered grid Lagrangian hydrodynamics following Von Neumann and Richtmyer \cite{vonneumann1950}. The adjoint solution is computed via the checkpointed back-propagation described above. To find the stability characteristics of the adjoint solution, the gradients computed via the adjoint solution were compared to those computed with finite difference (where each variable was perturbed individually to build up a Jacobian). This analysis was performed with a batch (10 members) of random initial conditions ($\theta_\rho$, $\theta_P$) as well as those with initially uniform pressure (as in an implosion problem). Results are shown in \cref{fig:gradient_stability_analysis}. From this we were able to conclude that the use of a smoothed version of von Neumann artificial viscosity (see \cref{eqn:smoothedQ}) was beneficial, when operated at $\lesssim$ 0.1 of the CFL stable forward time step.

Proceeding from the gradient stability analysis, optimisation of the Sod shock inverse design problem was performed using canonical gradient descent:
\begin{align}
    \theta_{n+1} = \theta_n - \alpha_n \frac{\partial L}{\partial \theta}
\end{align}
where $n$ is the index of the training ``epoch". In addition, other hyperparameters (namely $\lambda$ and VNR artificial viscosity coefficients) were varied throughout training. These were started at high values (pushing towards smoother solutions) and decayed towards lower values to allow the steep and discontinuous initial conditions to be converged upon.

\begin{figure}[h]
\centering
\includegraphics*[width=0.95\columnwidth]{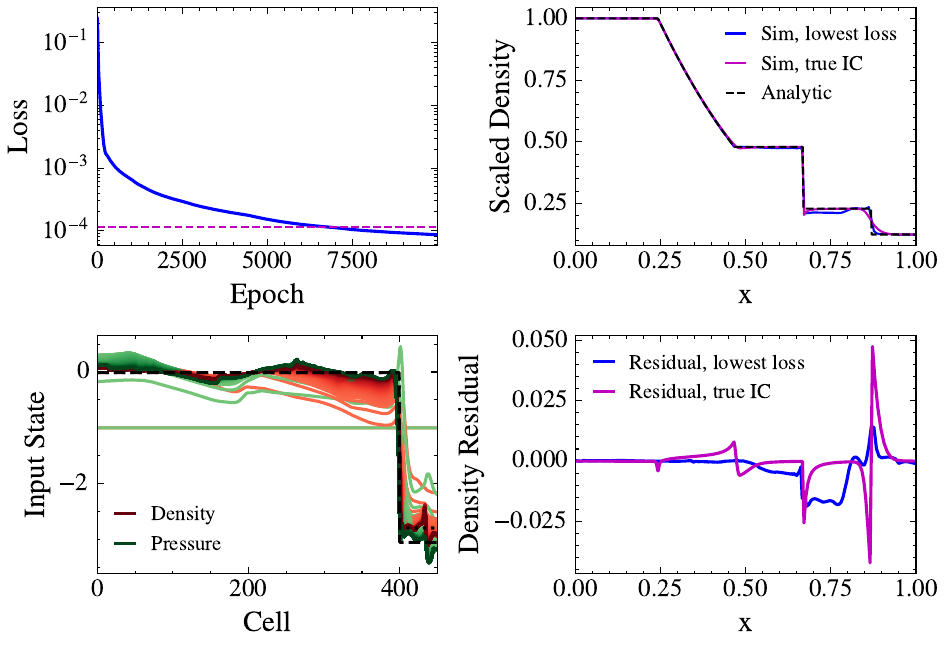}
\caption{(Top left) Distance (or loss) between reference and inverse design solution as a function of training step or epoch. The horizontal magenta line is the loss if one uses the true initial conditions as input to the numerical solver. (Top right) Plot of density comparing reference, final inverse design and true initial conditions results. (Bottom left) Evolution of the initial conditions within the inverse design, the opacity of the lines is used to represent that epoch. Note the additional perturbations about the true initial conditions shown as black dashed lines. (Bottom right) The residual between the numerical solutions and the reference solution.}
\label{fig:SodShockOptim}
\end{figure}

The results of the optimisation are shown in \cref{fig:SodShockOptim}. It is found that a set of initial conditions can be found which are broadly consistent with the true Sod Shock initial conditions, but with a lower mean squared error than a calculation performed with the true initial conditions. The optimisation scheme has therefore found a set of conditions which compensate for the deficiencies (numerical diffusion and other discretisation errors) in the numerical forward solution method. In particular, we note that the shock at $x \sim 0.9$ is sharper in the optimised conditions case. This is caused by the addition of density and pressure oscillations about the discontinuous initial conditions which launch additional waves.

In summary, we show that AD-derived parameter gradients can be used to optimise a high dimensional (900 parameters) design problem through the solution of compressible hydrodynamics. A careful hyper-parameter tuning scheme is required to achieve smooth training and stable adjoint solutions. Finally, as expected the final solution quality is tied to the accuracy of the forward solution scheme.

\section{Inverse pulse design for laser direct drive}\label{LDD_design}

In the following we explore inverse, gradient-based design of a laser direct drive implosion using 25 kJ of laser energy. The laser pulse will be parameterised with a large number ($\mathcal{O}(100s)$) of parameters, making it difficult and prohibitively expensive to optimise using popular gradient-free methods such as Bayesian optimisation. We also explore varying target parameters (fuel mass and outer radius) to find the optimal combination of laser pulse and target geometry.

An objective function is used to quantify progress towards the optimal design, where the objective function reaches an optimum. Ultimately, the goal of ICF is to optimise fusion gain, however for sub-scale (i.e. designs without sufficient energy to achieve ignition) implosions, which cannot ignite, a metric for proximity to ignition is required. One such ignition metric is $\chi_{no \ \alpha}$ \cite{Betti2015,Christopherson2020}, which depends on fusion yield, areal density ($\rho R$) and fuel mass. We define our objective function as:
\begin{align}
    O(\theta) &=  - \chi_{\mathrm{no} \ \alpha}(L(\theta)) + R(\theta) \ , \label{O_orig} \\
    \chi_{\mathrm{no} \ \alpha} &= \left(\frac{\rho R}{1 \mathrm{g/cm}^2}\right)^{0.61} \left(0.12\frac{Y_{DT}}{10^{16}}\frac{1 \ \mathrm{mg}}{M_{\mathrm{stag}}}\right)^{0.34} \label{chi_orig}
\end{align}
where $\theta$ are the parameters of the parameterised laser pulse, $R$ is a regularising term which steers the optimiser away from unphysical laser pulses or controls stability, $L(\theta)$ is the differentiable implosion code which predicts the evolution of implosion given the trainable laser parameters, $Y_{DT}$ is the DT yield, and $M_{\mathrm{stag}}$ is the stagnated mass. Within the objective function, the performance metric is negated because gradient-descent methods minimise the objective; the regularisation term is added to penalise undesirable pulses.

In design for ICF, the optimal pulse shape depends strongly on the target geometry. In the following, we aim to optimise the pulse, given the target parameters. We will separate the pulse and target design into a hierarchy of optimisation problems. This is performed for a number of reasons. Firstly, the dimensionality of target parameters is typically low, in our case it is two-dimensional with only ice mass and outer radius as parameters. We can therefore exploit batching of laser pulse optimisation runs in these dimensions, making use of the available compute by spreading the work across processors. Secondly, varying target parameters (such as ice mass and outer radius) can involve non-differentiable Lagrangian zoning calculations which would prevent gradient-based optimisation. In total, this would then lead to a hybrid optimisation approach where laser parameters are optimised (for fixed target parameters) via gradient-based methods. Then target parameters would be optimised on top of this via gradient-free optimisation such as those already used widely in the field\cite{gopalaswamy2025automated,Hatfield2019,Gopalaswamy2021,Ejaz2024,bronner2025particle,Wang2024,Spears2025,Humbird2019,vazirani2021coupling,vazirani2024bayesian,vazirani2025comparison}. In this work we will focus on the novel part of this optimisation scheme, the gradient-based optimisation of the pulse parameters. 

\subsubsection{Bounded interpolated pulse}

Constraints must be included into the functional form of the laser pulse such that key requirements are enforced by construction. Chiefly, these are that there is a fixed energy within the pulse and that there is an upper bound to laser power (set by both laser plasma instability risk and driver capability). From an optimiser perspective, it is also attractive to have the trainable parameters to not be restricted, and to have magnitudes of order unity. One of the simplest, but still generic, functions to respect these criteria is a linearly interpolated pulse where the trainable parameters are passed through a non-linear transform from $\mathcal{R}$ to $[0,P_{peak}]$. Mathematically, we define:
\begin{align}
    P(\theta ; t) &= \mathrm{Interpolate} \left(\hat{P}(\theta), \hat{t}(\theta); t\right) \ , \\
    \hat{P}(\theta) &= P_{peak} \cdot \frac{1}{2}\left(\tanh{\theta}+1 \right) \ , \\
    \hat{t}(\theta) &= \frac{E_{L}}{\mathrm{Trapezoid Rule}(\hat{P}, \hat{\tau})} \cdot \hat{\tau} \ , \\
    \hat{\tau} &= \mathrm{LinSpace}(0,1,\mathrm{len}(\theta)+1)
\end{align}
where $P_{peak}$ is peak power (25 TW), $E_L$ is the total laser energy (25 kJ), $\theta$ are the (500) trainable parameters that define the pulse shape. To convert this incident laser power into an intensity (and therefore ablation pressure), we use the initial hard sphere illumination area and account for blow-by of a super-gaussian beam:
\begin{align}
    I(t) = \frac{P(t)}{A_0} \cdot \mathrm{erf}\left(\frac{A_0}{A_{\mathrm{beam}}}\right) \ ,
\end{align}
where $A_{\mathrm{beam}} = \pi (358 \ \mathrm{um})^2$. To ensure we only consider smooth variations in the laser pulse, we define the following smoothness regularising term to the objective function:
\begin{equation}
    R(\theta) = \lambda \sum_{i=1,N-2} \left(\theta_{i-1}+\theta_{i+1}-2\theta_{i}\right)^2 
\end{equation}
where $\lambda$ is the smoothness hyperparameter which we will vary throughout the optimisation process. This term penalises high curvature in the laser pulse, therefore ensuring smoothness without penalising first-order gradients.

Optimisation was performed using the limited-memory Broyden–Fletcher–Goldfarb–Shanno bounded (L-BFGS-B) algorithm. This is a quasi-Newton method where an approximation of the inverse Hessian is updated using previous Jacobian and function values. The inverse Hessian is used to find search directions along which line-search steps are taken which only require function evaluations. It was found that the Jacobian evaluations (via AD-adjoint methods) of the differentiable implosion code were more expensive than pure function evaluations. Therefore, L-BFGS proved to be efficient for gradient-based optimisation as a mixture of Jacobian and pure function evaluations could be used. The bounds set on the trainable parameters were $5 > \theta > -5$, corresponding to maximum and minimum laser powers of 24.999 and 0.001 TW respectively.

\begin{figure}[h]
\centering
\includegraphics*[width=0.99\columnwidth]{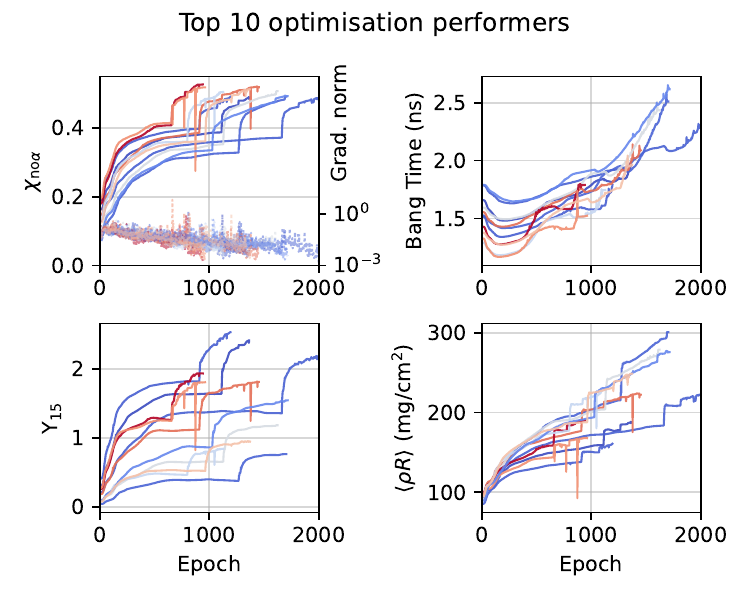}
\caption{Figure showing the progression of performance as a function of epoch/evaluations for the top 10 performers in the implosion design study. Each optimisation shows a rapid rise in $\chi_{\mathrm{no} \ \alpha}$ part way through the optimisation, this corresponds to the drop of initial temperature from 1 eV to 0.1 eV which leads to much improved designs, but comes at the cost of increased computational cost. (Top left) $\chi_{\mathrm{no} \ \alpha}$. Also on the left y-axis shown with the dotted lines are the L2 norms of the objective function gradients which exhibit large levels of noise. (Top right) Time of peak neutron production. (Bottom left) DT neutron yield $\times 10^{15}$. (Bottom right) Burn averaged fuel areal density. Lines coloured by peak $\chi_{\mathrm{no} \ \alpha}$ value achieved.}
\label{fig:Loss}
\end{figure}

As in the Sod shock tube example, to ensure stable and efficient optimisation, additional hyperparameters were varied throughout optimisation. These included: initial fuel temperature ($T_0$), regularisation parameter ($\lambda$), number of time steps ($N_t$), number of radial zones ($N_r$) and simulation duration ($t_{\mathrm{stop}}$). A piecewise constant schedule was devised where the schedule was updated once the L-BFGS optimisation had reached convergence with a given hyper-parameter set. Starting with an initially high $T_0$ and $\lambda$ allowed for a small number of numerical time steps needed to obtain stable gradient values, thus speeding up the initial optimisation. By incrementally dropping $T_0$ and $\lambda$ and increasing $N_r$, lower adiabat implosions were possible as well as sharper variation in pulse shape. However, these hyper-parameter values also required more $N_t$ in order to stabilise the adjoint solve.

Following the above methodology, batches ($\sim 50$ members) of gradient-based optimisations were launched with varied target parameters. Sobol sampling was used to perform quasi-random sampling of outer fuel radius (250-550 um) and fuel ice masses (5-20 ug). For each of these target geometries, an optimisation run was initialised with a laser pulse of the form given by a linear ramp up to 12.5 TW and back down, in transformed parameters $\theta$. Total convergence (through all of the hyper-parameter schedule) was commonly achieved with $\mathcal{O}(1000)$ epochs in the optimisation, as shown for the top 10 performers in \cref{fig:Loss}. This is notably more simulation efficient than gradient-free methods that require 100s-1000s of simulations to optimise 1s-10s of parameters\cite{gopalaswamy2025automated,Crilly2025}.  However, the differentiable method presented difficulties in adjoint stability as discussed above. Combined forward and adjoint simulations are more expensive ($\sim$ 5x) than the corresponding forward only, approximately 10-15 mins vs 2-3 mins on 4 CPUs. This however is considerably cheaper than performing finite difference estimation of the gradients, which would require, at least, an additional forward simulation for each design parameter, i.e. 500x more expensive.

\begin{figure}[h]
\centering
\includegraphics*[width=0.99\columnwidth]{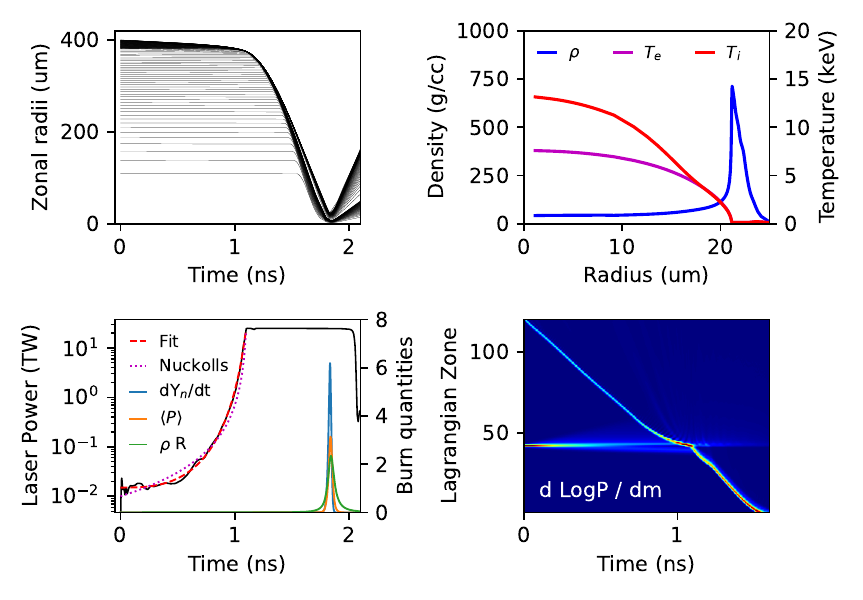}
\caption{Figure showing results from highest performance simulation in the optimisation campaign. (Top left) Zone radii as a function of time. (Top right) Density and temperature at the peak of neutron production. (Bottom left) Plot showing laser power against time, including a fit to the rise using Gopalaswamy \textit{et al.} function\cite{gopalaswamy2025automated} as well as Nuckolls isoentropic drive\cite{Nuckolls1972}. Also shown are the yield rate ($dY_n/dt$ in units of $10^{16}$ n/ns), burn-averaged pressure ($\langle P \rangle$ in units of 100 Gbar) and areal density ($\rho R$ in units of 100 mg/cm$^2$). (Bottom right) Logarithmic derivative of pressure, showing the single pulse shock as well as the rarefaction of the fuel shell caused by the initial pre-heat.}
\label{fig:OptimSim}
\end{figure}

The optimal design amongst this optimisation campaign was found for an outer fuel radius of 400 um and stagnated mass of 7.76 ug, achieving a $\chi_{\mathrm{no} \ \alpha}$ of 0.527, DT yield of 1.93 $\times 10^{15}$, and burn-averaged and bang-time stagnated fuel $\rho R$ of 190 and 221 mg/cm$^2$ respectively. The simulation results are summarised in \cref{fig:OptimSim}. The optimised laser pulse shows a clear ramp to peak power which does not launch additional strong shocks. The rise of the pulse was fit using the functional form given in Gopalaswamy \textit{et al.}\cite{gopalaswamy2025automated} - with best fit parameters of $\beta_1$ = 5.4, $\beta_2$ = 2.9 and $t_{\mathrm{rise}}$ = 1.1 ns, and fixed parameters $P_{0,\mathrm{laser}}$ = 0.015 TW and $P_{\mathrm{peak, laser}}$ = 25 TW:
\begin{align}
    P_{\mathrm{laser}} &= \frac{P_{0,\mathrm{laser}}}{\left[1 - \left(\frac{t}{t_0}\right)^{\beta_2}\right]^{\beta_1}} \ , \\
    t_0 &= t_{\mathrm{rise}} \cdot \left[1 - \left(\frac{P_{\mathrm{peak, laser}}}{P_{0,\mathrm{laser}}}\right)^{-\frac{1}{\beta_1 }}\right]^{-\frac{1}{\beta_2}} \ .  
\end{align}
This energy drive can be compared to the theoretical isoentropic compression of solid DT from Nuckolls\cite{Nuckolls1972}, which has $\beta_2$ = 1 and $\beta_1 = \frac{3 \gamma}{\gamma + 1} \left(=\frac{15}{8}\right)$, which is also plotted in \cref{fig:OptimSim}. From this comparison we can see that the gradient-based optimisation of Lawson parameter alone has arrived at a laser pulse which resembles literature results for the optimal spherical compression of DT shells.  The optimised drive allows the target to reach a very high implosion velocity of 645 km/s at a low adiabat approaching Fermi degeneracy ($\lesssim$ 1.5) and high IFAR (peak of 300 but one should note that we only simulate the stagnated fuel so underestimate the full shell thickness considerably). As noted above, for numerical stability reasons this simulation begins at a temperature of 0.1 eV which corresponds to $\sim$ 60 mJ of fuel preheat, using FEOS\cite{FEOS1,QEOS}. 

While the performance metrics for this OMEGA scale implosion are high (and optimistic), the design achieves stagnation and burn before the end of the laser pulse, suggesting wasted energy and a not fully optimised design. To a degree this can be attributed to the way the pressure drive is applied given the laser power. Coast is not sustainable in this simplified implosion model as once the laser power turns off the fuel instantly begins to rapidly rarefy into vacuum pressure. Therefore, the drive must be maintained during stagnation in the absence of the free-falling shell inertia. That said, there is still potential to improve upon the design (also shown in \cref{fig:Loss} although convergence has slowed significantly), here we present the key result showing that an isoentropic rise to peak power is found given no prior. Further optimisation may benefit from removing or augmenting the regularisation term (as it can introduce and promote oscillations in the pulse) and reduced time step to improve adjoint stability as gradient values became increasingly noisy approaching the optimum. This prompts further exploration of high dimensional pulse parameterisations which avoid pathological optimisation artifacts and issues.

This study shows that differentiable design can be used to optimise implosions with 100s of design parameters. While limited in choice of design parameters (laser pulse only) and fidelity of the physics, it stands as a proof of principle of this technique as a method for inertial fusion target design. In the following section, we explore neural networks as a method to parameterise the pulse and show benefits in terms of exploration and speed of convergence - although we do not aim to replicate the full design campaign performed for the bounded interpolated pulse presented in this section.

\subsubsection{Neural network pulse}
The laser pulse can also be parameterised by a neural network (NN), inspired by neural reparametrisation for structural design \cite{hoyer_neural_2019,Joglekar2024}. The pulse power as a function of time is:

\begin{equation}
	P(\theta ; t) = \mathrm{MLP}_\theta \left(\gamma \left( t \right) \right) \cdot P_{peak}
\end{equation}

where $\theta$ are the learnable weights of the multi-layer perceptron (MLP), $t \in [0, 1]$ the normalized time coordinate, and $\gamma$ the temporal encoding. The MLP uses rectified linear unit (ReLU) activations and varying depth $d$ and width $w$. A final \texttt{sigmoid} nonlinearity constrains the outputs between 0 and 1 in a differentiable manner. To mitigate spectral bias (the tendency for NNs to learn low-frequency features much faster than high-frequency ones), the time coordinate is transformed into a vector of Fourier features \cite{tancik_fourier_2020, wang_eigenvector_2021}, identical to commonly used positional encodings \cite{vaswani_attention_2017, mildenhall_nerf_2020}. Given the number of Fourier octaves $K$, the features are
\begin{equation}
    \omega_k = 2^k \pi, \qquad k = 0, 1, \ldots, K-1.
\end{equation}
The temporal encoding $\gamma : \mathbb{R} \to \mathbb{R}^{2K+1}$ is then
\begin{align}
    \gamma(t) &= \bigl(t,\; \sin(\omega_0 t),\; \cos(\omega_0 t),\; \ldots,\; \nonumber \\
    &\sin(\omega_{K-1} t),\; \cos(\omega_{K-1} t)\bigr).
\end{align}

This temporal encoding is found to be necessary to learn structure at multiple timescales, with $K=4$ being sufficient.

Total laser energy is constrained via modifying the loss (objective) function from eq. \ref{O_orig}. Using the same definition as $\chi_{\mathrm{no} \ \alpha}$ (eq. \ref{chi_orig}), the loss used here is,
\begin{equation}
	O(\theta) =  - \log \left( \chi_{\mathrm{no} \ \alpha}(L(\theta)) \right) + R(\theta) \ ,
\end{equation}
Using a $\log$ loss function appears to improve gradient stability when using this NN parameterisation.
Given the pulse energy, integrated using the trapezoid rule,
\begin{equation}
    E(\theta) = \int_0^{t_{\max}} P(\theta;t)\, dt,
\end{equation}
the energy penalty (regularisation term) is then
\begin{equation}
    R(\theta) = \lambda \left( \frac{\max(0,\, E - E_{\max})}{E_{\max}} \right)^2,
\end{equation}
where $E_{\max}$ is the 25 kJ energy budget and $\lambda$ the penalty strength. Constraining the total energy via a regularisation term was chosen over pulse clipping for smoothness of the loss landscape, to maintain nonzero gradients throughout the entire pulse, and to permit exploration of optima that may be otherwise inaccessible without temporarily exceeding this energy constraint.

In this optimisation we keep the shell configuration fixed, using an ice mass of 15 ${\mu}$g and an outer shell radius of 490 $\mu$m. The initial temperature is also held constant at 1 eV across simulation iterations, as mentioned in previous sections this relaxed the timestep constraint for the adjoint solve. The timestep is fixed at 100,000 steps per simulation iteration for 3.5 ns ($\Delta t = 3.5$ fs) for simulator stability across a variety of pulse shapes. Total simulation (forward and adjoint) run time per step was approximately 4.5 minutes. This shows there is minimal runtime cost in a vast increase in pulse parameters, from $\mathcal{O}(100)$ in the previous section to $\mathcal{O}(100,000)$ for the NN pulse. 
\begin{figure}[b]
	\centering
	\includegraphics[width=0.99\columnwidth]{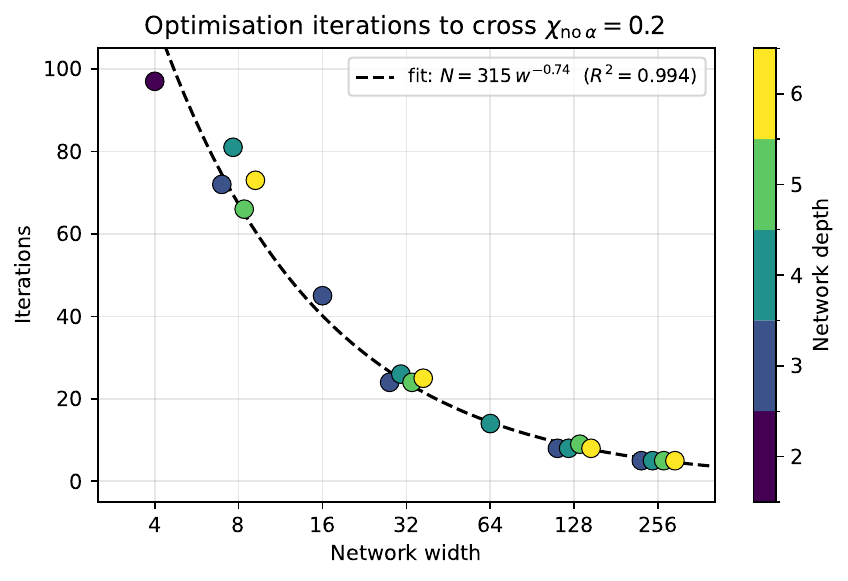}
	\caption{\label{fig:nn-lawson-depth-width}Number of optimisation iterations needed to exceed $\chi_{\mathrm{no} \ \alpha} = 0.2$ as a function of the MLP layer width and depth. Runs with varying depth are spread horizontally for clarity. Increased capacity by increasing layer width improves the pace at which the network explores the loss surface, and the iterations scale with width following a power law (exponent of -0.74).}
\end{figure}
Optimisations were performed in two stages: an exploratory phase and a fine-tuning phase. The weights of the MLP were tuned using the \texttt{Adam} optimiser\cite{kingma2014adam}. In the exploratory phase, a learning rate of $3 \times 10^{-3}$ with $\beta_1=0.9$ and $\beta_2=0.999$ was used. A cosine decay learning rate schedule was used over 500 iterations, though the effect of this schedule is relatively small because of the $\mathcal{O}(100)$ iterations needed. The regulariser weight $\lambda$ was set to 0.1.

The NNs were set with a width of 256 and a depth of 6, totaling $\approx$ 331,000 parameters, for rapid exploration. To find the architecture which took the fewest epochs to learn, the effect of width and depth of the networks was investigated. Increasing the parameter count of the neural network dramatically increases the rate at which good pulses are found, as seen in fig \ref{fig:nn-lawson-depth-width}, which compares NNs with widths from 4 to 256 and depths from 2 to 6. Here we compare the number of iterations before $\chi_{\mathrm{no} \ \alpha}$ improves rather than the achieved optimum, as subsequent adjustments to the optimiser hyperparameters can then be used to improve convergence to the optimum. The ability of an NN to learn complicated curves depends polynomially on the width and exponentially on the depth of the network \cite{telgarsky_benefits_2016}. In practice, increasing the width of the network always improves the speed of training. The relationship with depth is not so clear. The run time of the optimisation is dominated by the computational cost of the simulator itself, so using large NNs for reparameterisation may be ideal. 

\begin{figure}[t]
	\centering
	\includegraphics[width=0.99\columnwidth]{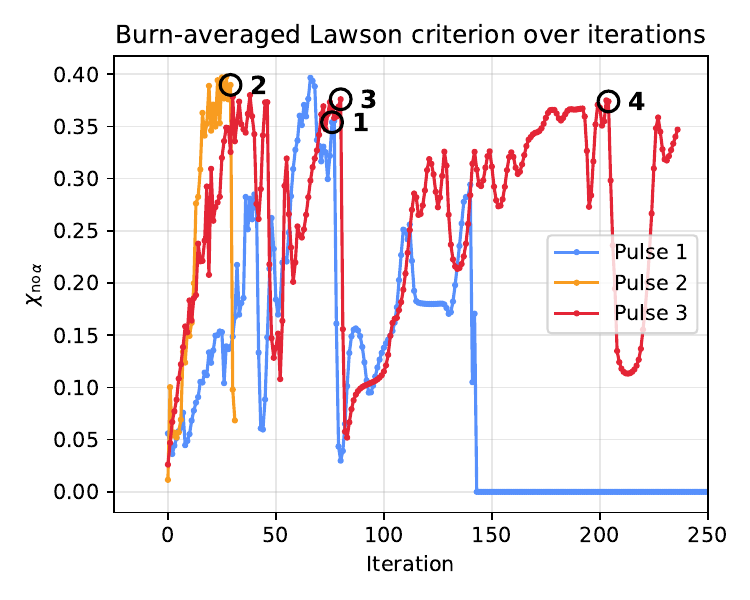}
	\caption{\label{fig:nn-lawson}Figure shows three well-performing optimisation trajectories initialized on different pulse shapes. The neural-network parameterisation of the laser pulse leads to rapid exploration of parameter space, quickly finding high-performing pulses but often overshooting the optimum. Four high-performing pulses of interest are circled, shown in fig. \ref{fig:nn-pulse-highlight}.}
\end{figure}

These NNs were pretrained using four differing pulse shapes: a single-picket followed by an isentropic rise and flat top pulse (similar to Gopalaswamy \textit{et al.}\cite{gopalaswamy2025automated} and Crilly \textit{et al.}\cite{Crilly2025}), Gaussian pulse train, rectangular pulse train, and combinations of sin, cos, and polynomials that were then clipped between 0 and 1 - then scaled to peak power of 25 TW. The optimisation trajectory was found to be sensitive to the initial pretrained configuration. The precise effect of the choice of pretrained pulse family is not clear, but the diversity of initial conditions aids exploration of pulse parameter space in parallel. If the network is not pretrained, the resulting pulse is a random, smooth curve that is increasingly flattened with increased depth and width of the MLP with an offset. In addition, pretraining is an important diagnostic because learning a pulse shape makes obvious the inductive bias of the networks: lower-capacity NNs are unable to fit particular types of pulses, particularly ones with sharp features. If an NN cannot learn a pretrained pulse, it will likely struggle to learn that pulse when embedded with a differentiable simulator in the loop.

Examples of three high-performing ($\chi_{\mathrm{no} \ \alpha} > 0.3$ at least once) optimisation trajectories can be seen in fig. \ref{fig:nn-lawson}. Three other high-performing trajectories are not shown out of twelve total for clarity. Using an NN-based pulse representation enables rapid exploration of parameter space, hitting Lawson criteria of $~\sim 0.4$ in less than 50 iterations in some cases. However, the initial optimiser settings appear to cause overshoot of these optima, possibly because of momentum terms (essentially gradient memory) in the \texttt{Adam} optimiser. Future work will look to stabilise these training trajectories for neural parameterisations.

\begin{figure}[t]
	\centering
	\includegraphics[width=0.99\columnwidth]{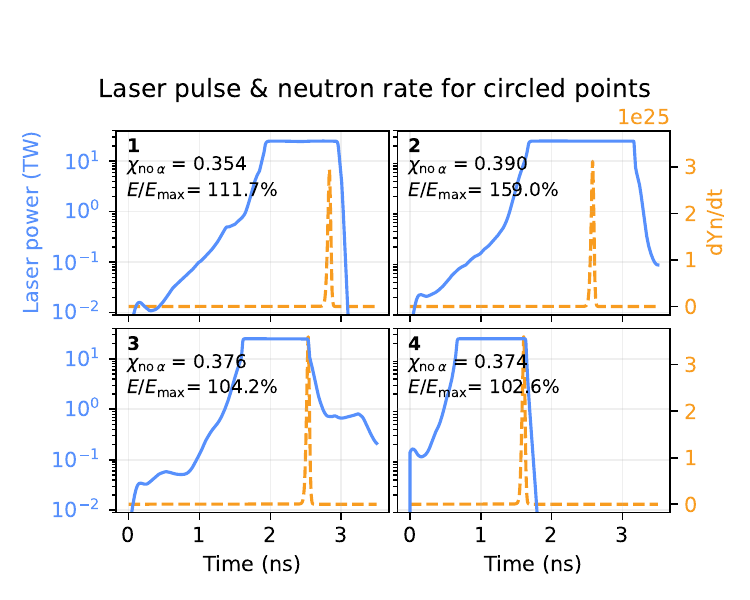}
	\caption{\label{fig:nn-pulse-highlight}Laser pulses (log scale) from the circled pulses in fig. \ref{fig:nn-lawson}. All pulses achieved similar $\chi_{\mathrm{no} \ \alpha}$ with relatively similar shapes, a picket and/or foot followed by a rise to peak power with a flat top. Each pulse also used excess energy.}
\end{figure}

\begin{figure}[h]
	\centering
	\includegraphics[width=0.99\columnwidth]{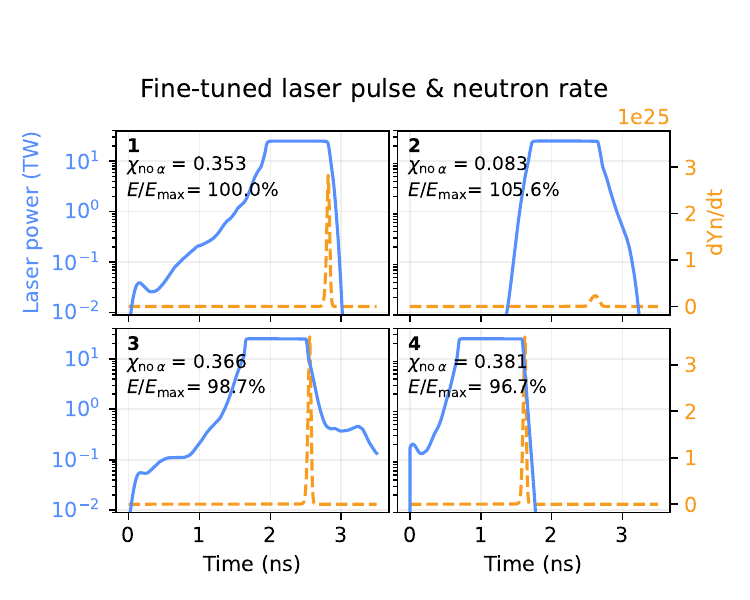}
	\caption{\label{fig:nn-pulse-highlight-resume}Laser pulses from fig. \ref{fig:nn-pulse-highlight} fine-tuned using a smaller learning rate and an aggressive energy regularisation ($\lambda$) schedule. Fine tuning improved adherence to the total energy constraint but the pulse shapes remained similar. Notably, pulse number 2 (top right) had an extreme energy constraint violation that needed rectification before the pulse could be improved. The enforcement of the penalty term removed the foot and rise of the pulse to compensate, therefore leading to reduced burn performance.}
\end{figure}

Four high-performing pulses, circled in fig. \ref{fig:nn-lawson}, can be seen in fig. \ref{fig:nn-pulse-highlight} with the neutron yield rate. All pulses have a similar shape, with variation at the foot and tail of the pulses. The regularisation term in these initial pulses is too weak to sufficiently constrain the total energy. All four of the shown pulses exceed the energy constraint by at least 2.6\%, with one pulse exceeding by 59\%. In this second phase of optimisation, these pulses are fine-tuned to more strictly enforce this energy constraint by loading the network weights and restarting the optimisation. The fine-tuning steps use a smaller learning rate of $1\times 10^{-4}$, cosine-decaying over 300 iterations, and ramp $\lambda$ from 0.1 to 10 over 100 epochs. This reloading of weights also effectively resets the \texttt{Adam} optimiser state; the momentum terms from the previous iterations are not preserved. The resulting pulses can be seen in fig. \ref{fig:nn-pulse-highlight-resume}. Shifting the weight towards more strictly enforcing the energy term does reduce the energy violation, however, as can be seen for pulse 2, this does not guarantee maintaining performance. This points to using an exploratory batching approach where many NN pulses are optimised and fine-tuned in parallel, allowing the best candidate to be selected at the end of the process. 

In conclusion, the neural reparametrisation of the laser pulse shape can dramatically improve the speed of finding laser drive pulses with high performance metrics. However, fine-tuning of the highest-performing pulses may be required to mitigate overshoot and respect constraints. The effect of the NN architecture, width, depth, capacity, activations, and so on, are unclear, but the results presented here encourage use of neural reparametrisation in future gradient-based optimisation studies to increase the speed of convergence for arbitrary pulse shapes, allowing more rapid exploration.

\section{Conclusions}

In this work we present the novel application of differentiable programming techniques to implosion modelling and design in the context of inertial confinement fusion. By using JAX\cite{jax} and associated libraries, a fully differentiable Lagrangian hydrodynamics model was developed and used to investigate the solution to the adjoint equations, i.e. obtaining gradients with respect to some scalar objective function. These gradients obtained via Automatic Differentiation (AD) were then used to perform efficient, gradient-based optimisation, solving an inverse design problem. This is in contrast with current automated design methodologies in inertial confinement fusion\cite{gopalaswamy2025automated,Hatfield2019,Gopalaswamy2021,Ejaz2024,bronner2025particle,Wang2024,Spears2025,Humbird2019,vazirani2021coupling,vazirani2024bayesian,vazirani2025comparison}, which rely on existing (non-differentiable) simulation codes and gradient-free optimisation methods, such as Bayesian optimisation\cite{Snoek2012}. 

Firstly, we used the Sod shock tube as a simple test problem to showcase successful gradient based optimisation via differentiable simulation. A number of interesting findings were established through this study, with a key result being that the adjoint system found via AD was more numerically stiff than the forward system. Some reduction in numerical stiffness was possible by smoothing the artificial viscosity function. Still, this study found that an order of magnitude smaller time step was typically needed for a stable adjoint problem solution. This prompts further work to investigate the source of this numerical stiffness and possible improvements. This result also marks a current limitation of differentiable simulation methods for inverse design as the computational and memory cost of stably solving the AD-derived adjoint system can become prohibitive.

Secondly, we used the differentiable implosion code to optimise a high-dimensional direct-drive pulse design via an external pressure boundary condition\cite{HTL2001}. This involved the tuning of 500 trainable parameters to maximise the Lawson parameter as given by $\chi_{\mathrm{no }\alpha}$. The pulse was parameterised in a general manner, allowing a very wide variety of pulse shapes to be described. Various target geometries were sampled and individually optimised via the LBFGS-B algorithm using AD gradients. The best performing pulse and target design showed canonical features of optimal implosion design; a low adiabat, fast implosion driven by a near isoentropic rise to peak power followed by a flat top. The method still has key limitations, stemming from the reduced physical model and slowed final convergence, which set this work as a proof of principle rather than a fully capable design tool.

Finally, neural reparameterisation of the pulse shape enables rapid exploration of arbitrary pulse shapes with fewer epochs to near-optimum. The forward and adjoint runtime is only weakly dependent on the number of trainable parameters; in contrast to naive finite-differencing, AD enables $>$ 100,000x speedup. Increasing width of the network enables faster exploration, while Fourier features improve the types of curves the MLP can represent. Pretraining of the MLP on a pulse shape highlights the inductive biases present in the MLP, and provides a suitable initialisation for the network. Regularisation can be used to enforce laser energy requirements; we showed that a staged approach with fine-tuning preserved performance while improving adherence to the energy constraint for the majority of cases.

While this work shows promising results for AD-enabled inverse design, there are a number of avenues of future work to be explored with an aim to improve the simulation fidelity, optimisation strategy and adjoint robustness. For simulation fidelity, there are a number of physical phenomena not captured by the current model, most notably the laser ablation and corona. While including these models would improve the fidelity, it is the case that numerical models for ICF (especially 1D codes) are not predictive of experiment. A possible unique advantage of a differentiable code is the ability to embed trainable machine learning (ML) models within the physical system of partial differential equations. These can then be trained against high fidelity simulations or experiment to improve predictive capability. For non-differentiable codes, this ML model correction typically occurs external to the simulator, for example the statistical model trained on OMEGA experiments and LILAC simulations\cite{Gopalaswamy2019}. A recent exception to this is the work of Follett \textit{et al.}\cite{follett2025}, which included trainable, 3D degradation physics terms into LILAC\cite{Delettrez1987} with a smaller number of parameters. A differentiable version of this approach would allow for many more trainable parameters in the embedded models allowing for more complex functions to be learnt. For optimisation strategy, it is the case that the performance of gradient-based methods depends strongly on parameterisation and objective function, which has not been comprehensively explored in this work. By establishing a benchmark for ICF implosion design, improvements in the optimisation methods can be found empirically. Additionally, simulation bias correction terms (like the statistical model\cite{Gopalaswamy2019,Lees2025}  developed at LLE) can be included in the objective function such that more realistic designs can be optimised for, even when using a 1D simulator. Finally, adjoint robustness is a unique constraint of differentiable simulation and therefore not explored in previous work in the context of numerical models of ICF. Future work into the source of adjoint stiffness for canonical numerical schemes (such as staggered grid Lagrangian hydrodynamics) as well as less stiff methods is required. It may be the case that using AD to form the adjoint directly from the forward solution method is not ideal for all solvers\cite{Rackauckas} in an operator split code. Instead, custom adjoint solvers could be needed to ease the time stepping requirements, likely at the cost of gradient accuracy.

\section*{Acknowledgments}
This research received support through Schmidt Sciences, LLC, the International Atomic Energy Agency AI for Fusion CRP, the Imperial College Research Fellowship program, Pasteur Labs and the Institute for Simulation Intelligence (ISI), and IFE COLoR under U.S. Department of Energy Grant No. DE-SC002486. 

This research used resources of the Imperial College London RCS High Performance Computing systems and the National Energy Research Scientific Computing Center, a DOE Office of Science User Facility supported by the Office of Science of the U.S. Department of Energy under Contract No. DE-AC02-05CH11231 using NERSC award FES-ERCAP0026741. 

The authors declare no competing interests.

\section*{References}
\bibliography{MuCFRefs}

\end{document}